\documentclass[11pt]{article}
\usepackage{graphicx,epsf,amssymb}

\newcommand{\be}{\begin{equation}}
\newcommand{\ee}{\end{equation}}
\newcommand{\ba}{\begin{eqnarray}}
\newcommand{\ea}{\end{eqnarray}}

\newcommand{\bea}{\begin{eqnarray}}
\newcommand{\eea}{\end{eqnarray}}

\newcommand{\NeqFour}{{\cal N} =4}

\def\fig#1{fig.~{\ref{#1}}}

\def\NeqFour{{\cal N}=4}
\def\NeqOne{{\cal N}=1}

\def\Shift#1#2{{[#1,#2\rangle}}

\def\sandp#1.#2.#3{%
\left\langle\smash{#1}{\vphantom1}^{-}\right|{#2}%
\left|\smash{#3}{\vphantom1}^{+}\right\rangle}
\def\sandpp#1.#2.#3{%
\left\langle\smash{#1}{\vphantom1}^{+}\right|{#2}%
\left|\smash{#3}{\vphantom1}^{+}\right\rangle}
\def\sandmm#1.#2.#3{%
\left\langle\smash{#1}{\vphantom1}^{-}\right|{#2}%
\left|\smash{#3}{\vphantom1}^{-}\right\rangle}
\def\spab#1.#2.#3{\sandmm#1.#2.#3}
\def\spba#1.#2.#3{\sandpp#1.#2.#3}
\def\spaa#1.#2.#3.#4{\sandmp#1.{#2#3}.#4}
\def\spbb#1.#2.#3.#4{\sandpm#1.{#2#3}.#4}
\def\spa#1.#2{\langle#1\,#2\rangle}
\def\spb#1.#2{[#1\,#2]}
\def\spash#1.#2{\vphantom{\hat K}\spa{\smash{#1}}.{\smash{#2}}}
\def\spbsh#1.#2{\vphantom{\hat K}\spb{\smash{#1}}.{\smash{#2}}}
\def\lor#1.#2{\left(#1\,#2\right)}
\def\sand#1.#2.#3{%
\left\langle\smash{#1}{\vphantom1}^{-}\right|{#2}%
\left|\smash{#3}{\vphantom1}^{-}\right\rangle}
\def\sandpp#1.#2.#3{%
\left\langle\smash{#1}{\vphantom1}^{+}\right|{#2}%
\left|\smash{#3}{\vphantom1}^{+}\right\rangle}
\def\sandpm#1.#2.#3{%
\left\langle\smash{#1}{\vphantom1}^{+}\right|{#2}%
\left|\smash{#3}{\vphantom1}^{-}\right\rangle}
\def\sandmp#1.#2.#3{%
\left\langle\smash{#1}{\vphantom1}^{-}\right|{#2}%
\left|\smash{#3}{\vphantom1}^{+}\right\rangle}

\newbox\SlashedBox
\def\slashed#1{\setbox\SlashedBox=\hbox{#1}
\hbox to 0pt{\hbox to 1\wd\SlashedBox{\hfil/\hfil}\hss}#1}
\def\hboxtosizeof#1#2{\setbox\SlashedBox=\hbox{#1}
\hbox to 1\wd\SlashedBox{#2}}

\newbox\charbox
\newbox\slabox
\def\s#1{{      % Feynman slash
        \setbox\charbox=\hbox{$#1$}
        \setbox\slabox=\hbox{$/$}
        \dimen\charbox=\ht\slabox
        \advance\dimen\charbox by -\dp\slabox
        \advance\dimen\charbox by -\ht\charbox
        \advance\dimen\charbox by \dp\charbox
        \divide\dimen\charbox by 2
        \raise-\dimen\charbox\hbox to \wd\charbox{\hss/\hss}
        \llap{$#1$}
}}

\def\eqn#1{eq.~(\ref{#1})}

\def\e{\epsilon}

\def\tree{{\rm tree}}

\def\cg{c_\Gamma}

\def\sandp#1.#2.#3{%
\left\langle\smash{#1}{\vphantom1}^{+}\right|{#2}%
\left|\smash{#3}{\vphantom1}^{+}\right\rangle}

\def\Den#1#2 {\prod\limits_{k=#1}^{#2} \spa{k}.{(k+1)}}

\def\Res{\mathop{\rm Res}}
\def\tlambda{{\tilde\lambda}}

\def\Cuth{{\widehat {C}}}

\def\Vertex{R}

\def\PureCut{C}
\def\Res{\mathop{\rm Res}}

\def\Overlap{O}
\def\Inf{\mathop{\rm Inf}}

\newcommand{\Bmp}[1]{\langle #1\rangle}

\newcommand{\Asdef}{A_s}
\newcommand{\At}{A^{\tree}}

\newcommand{\hK}{\hat{K}}

\newbox\ourfigbox
\def\SizedFigureWithCaption#1#2#3{%
\setbox\ourfigbox
  \hbox{\hss\epsfxsize #1 \epsfbox{#2}\hss}
\hbox to \wd\ourfigbox{\vbox{\noindent\copy\ourfigbox\break
\vskip -6mm      \hbox to \wd\ourfigbox{\hss#3\hss}}}
}

\def\llongrightarrow{%
\relbar\mskip-0.5mu\joinrel\mskip-0.5mu\relbar
     \mskip-0.5mu\joinrel\longrightarrow}
\def\inlimit^#1{\buildrel#1\over\llongrightarrow}
\def\dash{\hbox{-\kern-.02em}}

\begin{document}

\begin{titlepage}

\begin{flushright}
%\TimeStamp

SPhT-T06/093. \\

\end{flushright}

\vskip 2.cm

\begin{center}
  \begin{Large} {\bf On-shell Recursion Relations for n-point
      QCD\footnote{Work in collaboration with C.~F.~Berger, Z.~Bern,
        L.~J.~Dixon \& D.~A.~Kosower.}\footnote{Talk given at the {\it
          7th Workshop On Continuous Advances In QCD, 11-14 May 2006,
          Minneapolis, Minnesota.}}  }

\vskip 1.cm

\end{Large}

\vskip 1.cm
{\large
Darren Forde.
} 

\vskip 0.5cm

{\it  Service de Physique Th\'eorique\footnote{Laboratory
   of the {\it Direction des Sciences de la Mati\`ere\/}
   of the {\it Commissariat \`a l'Energie Atomique\/} of France.},
   CEA--Saclay\\
          F--91191 Gif-sur-Yvette cedex, France }

\vskip 1cm

\begin{abstract}
  We present on the use of on-shell recursion relations. These can be
  used not only for calculating tree amplitudes, including those with
  masses, but also to compute analytically the missing rational terms
  of one-loop QCD amplitudes. Combined with the cut-containing pieces
  calculated using a unitarity approach complete one-loop QCD
  amplitudes can be derived. This approach is discussed in the context
  of the adjacent 2-minus all-multiplicity QCD gluon amplitude.
\end{abstract}

\end{center}

\vfill

\end{titlepage}

\section{Introduction}
\label{IntroSection}

The forthcoming experimental program at CERN's Large Hadron Collider
will place new demands on theoretical calculations.  In order to reach
the precision required by searches for and measurements of new
physics, these processes need to be computed to next-to-leading order
(NLO), which entails the computation of one-loop amplitudes.  These
are challenging calculations.  State-of-the-art Feynman-diagrammatic
computations have only recently reached six-point
amplitudes~\cite{Numerical} due to the large numbers of diagrams
involved.

Feynman diagram techniques are not the only method for performing
these needed one-loop contributions. Within the unitarity-based
method~\cite{UnitarityMethod,Allmultcutcomppart,MassiveUnitarity} and
related recent developments~\cite{ZFourPartons,BCFUnitarity}, one can
decompose one-loop colour-ordered gluonic QCD amplitudes into pieces
corresponding to $\NeqFour$, $\NeqOne$, and scalar contributions as
$A_n = A_n^{\NeqFour} - 4 A_n^{\NeqOne} + A_n^{\rm scalar}$. The
supersymmetric contributions can be computed by performing the cut
algebra strictly in four dimensions, with only the loop integrations
computed in $D=4-2\e$ dimensions.  Scalar-loop contributions require
that the cut algebra, and the corresponding tree amplitudes fed into
the unitarity machinery, also be computed in $D=4-2\e$
dimensions~\cite{MassiveUnitarity,RationalUnitarity,DDimExamples}. This
makes the computation of these pieces somewhat more difficult than in
the supersymmetric case and leads us to desire a more efficient
approach.

At one loop, computing a scalar loop in $D=4-2\e$ dimensions is
equivalent to computing a massive scalar loop in $D=4$ dimensions, and
then integrating over the mass with an appropriate weighting. The
computation of tree-level amplitudes with massive scalars is thus of
use in the unitarity method for computing massless loop amplitudes in
non-supersymmetric gauge theories.  On-shell recursion relations can
be applied to calculate the necessary tree
amplitudes~\cite{MassiveRecursion,FMassive}.  These relations extend
the tree-level on-shell recursion relations of Britto, Cachazo, Feng,
and Witten~\cite{BCFW}. The remarkable generality and simplicity of the
proof of these recursion relations, requiring only Cauchy's theorem
and a knowledge of the factorisation properties of the amplitudes, has
enabled widespread application at tree level~\cite{TreeRecurResults}
and even at loop
level~\cite{OnShellRecurrenceI,Bootstrap,Forde:2005hh}.

Using massive scalars, although more straightforward than the
unitarity method in $D=4-2\e$ dimensions, is still not the most
efficient applicable technique. More efficient still is an updated
version of the unitarity bootstrap technique~\cite{ZFourPartons}.  This
technique relies on first obtaining the cut-constructible parts of a
desired amplitude --- those terms containing polylogarithms,
logarithms, and associated $\pi^2$ terms --- via the unitarity method
in $D=4$. The missing rational terms that this process cannot capture
are then derived using one-loop on-shell recursion
relations~\cite{Bootstrap}.  This allows for a practical and systematic
construction of the rational terms of loop amplitudes.

\section{Recursive Bootstrap Approach}
\label{RecursionReviewSection}

Before describing the extension of the on-shell recursion relations to
loop processes we first give an overview of the tree level recursion
relations including their application to massive theories. In the
simplest case the recursion relations employ a parameter-dependent
`$[j,l\rangle$' shift of two of the external massless spinors, $j$ and
$l$, in an $n$-point process,
\begin{eqnarray}
[j,l\rangle: \qquad
&\tlambda_j &\rightarrow \tlambda_j - z\tlambda_l \,, \,\,\,\lambda_j\rightarrow\lambda_j\,, \nonumber\\
&\lambda_l &\rightarrow \lambda_l + z\lambda_j \,, \,\,\,\tilde{\lambda}_l\rightarrow\tilde{\lambda}_l\,,
\label{SpinorShift}
\end{eqnarray}
where $z$ is a complex parameter.  These shifted momentum then remain
massless, $k_j^2(z) = k_l^2(z) = 0$, and overall momentum conservation
is maintained.  Shifting massive particles is also possible although
more involved~\cite{MassiveRecursion,FMassive}. An on-shell amplitude
containing the momenta $k_j$ and $k_l$ then becomes
parameter-dependent as well.  Exploiting Cauchy's theorem to construct
the analytic tree level function $A(z)$ from its residues and assuming
that there is no contribution from the circular contour at infinity
allows us to solve for the physical amplitude $A(0)$,
\begin{equation}
A(0) = -\sum_{{\rm poles}\ \alpha} \Res_{z=z_\alpha}  \frac{A(z)}{z}\,.
\label{NoSurface}
\end{equation}

\begin{figure}[t]
\centerline{\epsfxsize 2 truein\epsfbox{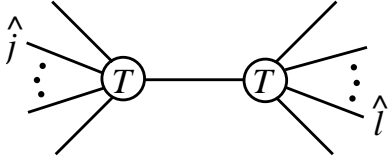}}
\caption{Schematic representation of a tree-level recursive
contribution to $A_n(0)$.  The labels `$T$' refer to tree vertices which are
on-shell amplitudes. The momenta $\hat{\jmath}$ and $\hat{l}$
are shifted, on-shell momenta. }
\label{TreeGenericFigure}
\end{figure}

The residues in \eqn{NoSurface} may be obtained using the generic
factorisation properties that any amplitude must
satisfy~\cite{TreeReview}. 
The propagator in any factorised channel where the shifted legs $j$
and $l$ lie on opposite sides of the pole, as depicted in
\fig{TreeGenericFigure}, will be of the form $1/(K^2-M^2-z\Bmp{j^-|\s
  K|l^-})$. Each pole therefore corresponds to a single factorised
channel and hence evaluating the residues of all such poles results in
an on-shell recurrence relation for $A(0)$ written schematically as
\begin{eqnarray}
A(0)=\sum_{\textrm{channels}}\sum_{h=\pm}
A(\ldots,\hat{j},\ldots,(-\hK)^{h})
\frac{i}{K^2-M^2}
A(\hK^{-h},\ldots,\hat{l},\ldots)\,.
\end{eqnarray}
In both amplitudes the momenta are all on-shell including the
intermediate momentum $\hK$, which can be massive (i.e. $\hK^2=M^2$).
Including massive external particles is therefore as straightforward
as using, where necessary, massive propagators and amplitudes with the
appropriate massive legs~\cite{MassiveRecursion,FMassive}.
%%%%%% FIGURE %%%%%%%%%%%
%%%%%%%%%%%%%%%%%

At loop level a number of new features arise.  In particular,
obtaining an on-shell recursion relation requires dealing with branch
cuts, spurious singularities, and in some cases, the treatment of
factorisation using complex momenta, which can differ from `ordinary'
factorisation using real momenta. An example of this, which applies
also at tree level, is the vanishing of all three-point vertices in
real momentum due to the constraints of momentum conservation. When
using complex momentum this is no longer the case as for complex
spinors $\overline{\lambda}\, \s \propto \,\tilde{\lambda}$, and so we
must now include three-point amplitudes in the recursion relations.
We must also contend with the possible appearance of double poles and
unreal poles in two-particle channels with like-helicity
gluons~\cite{OnShellRecurrenceI,Bootstrap}.

To set up a loop-level on-shell recursion we decompose the amplitude
into `pure-cut' and `rational' pieces, $ A_n(z) = \cg {}
\left[\PureCut_n(z) + \Vertex_n(z) \right]$.  The rational parts $R_n$
are defined by setting all logarithms, polylogarithms, and associated
$\pi^2$ terms to zero. It is then possible to show that the complete
amplitude at one-loop is given by~\cite{Bootstrap}
\be
A_n(0)
= \Inf A_n +  \cg {} \Biggl[ \Cuth_n(0) - \Inf\Cuth_n + R_n^D + O_n
 \Biggr]
\,,
\label{BasicFormula}
\ee
where $\Inf A_n$ is the potential contribution to the amplitude from
large $z$, $\Cuth_n(0)$ is the completed-cut contribution, which can
be calculated using unitarity based methods, $\Inf\Cuth_n$ is the
potential large-$z$ spurious behaviour of the completed cut, which
must be subtracted off, $R_n^D$ are the recursive diagram
contributions derived using an on-shell recursion relation, and the
`overlap' terms $O_n$ remove double counting between the recursive
diagrams and the rational terms that were added to complete the
cuts.\footnote{for a more detailed account of this see C.~F.~Berger's
  conference proceedings~\cite{CarolaProcCAQCD06}.}

\section{Solving recursion relations and all-multiplicity amplitudes}
\label{AllnSection}

Our basic stratagem to derive the complete one-loop amplitude is
therefore to calculate the cut-constructible pieces and then using
these construct the overlap terms. The remaining $R_n^D$ and
$\textrm{Inf}\;A_n$ terms are then calculated using an on-shell
recursion relation. Usually we will know the form of an amplitude
only up to a certain number of negative helicity legs (for a mostly
plus amplitude) and desire the form of the amplitude with one more
negative helicity leg. On constructing a recursion relation though we
will find that in some cases the recursion will contain an amplitude
with the same number of negative helicity legs, though fewer positive.
This is potentially problematic. For example consider the
all-multiplicity one-loop amplitude
$A^{\textrm{scalar}}_n(1^-,2^-,3^+,\ldots,n^+)$ the rational terms of
this amplitude $R_n(1,2)$ will be given after a $[1,2\rangle$ shift by
\begin{eqnarray}
R_n(1,2) &=&
\Asdef(1^-,2^-,3^+,...,n^+)
\nonumber\\
&&
  +R_{n-1}(\hat{1}^-,\hat{K}_{23}^-,4^+,...,n^+)\frac{1}{K^2_{23}}
A^\tree_3((-\hat{K}_{23})^+,\hat{2}^-,3^+)\,.
\label{eq:full_rec_rel}
\end{eqnarray}
Contained in $\Asdef$ are terms which are already known; the one-loop
amplitudes with one negative-helicity leg (which are completely
rational) and the tree amplitudes that multiply them.  The second term
contains $R_{n-1}$ which is the amplitude we are solving for but with
one less positive helicity leg.

\begin{figure}[t]
\centerline{\epsfxsize 3.75 truein\epsfbox{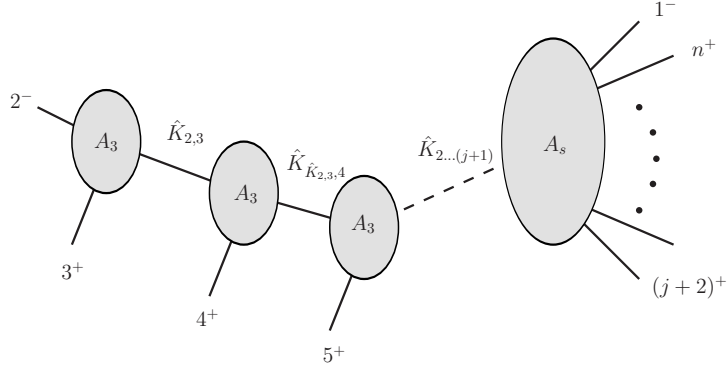}}
\caption{A contribution to the ``unwinding'', the string of $A_3$
  terms simplifies to
  $A_{j+1}(1^-,2^-,3^+,\ldots,(j+1)^+)/\Bmp{1\hK_{2\ldots{j+1}}}^2$.}
\label{figure:unwinding_result}
\end{figure}

Our tactic to solve \eqn{eq:full_rec_rel} for $R_n(1,2)$ is to insert
the left-hand side of \eqn{eq:full_rec_rel} into the right-hand side
of \eqn{eq:full_rec_rel} repeatedly. At each insertion we find that
our desired amplitude $R(1,2)$ appears on the right-hand side with one
fewer positive-helicity leg, and multiplied by one more three-point
gluon vertex and propagator, $\At_{3}/K^2$. This `unwinding' of the
amplitude continues until we have reduced the right-hand side of $R_n$
(\eqn{eq:full_rec_rel}) down to, in this case, $R_4=0$ and a sum of
terms that contain only known quantities (e.g. $\Asdef$ and overlap
terms $O$) multiplied by strings of $\At_3$ vertices, a contributing
example is shown in figure~\ref{figure:unwinding_result}.
%%%%%% FIGURE %%%%%%%%%%%
%%%%%%%%%%%%%%%%%

At each step of the unwinding we must choose new shifted momenta.  We
always choose to shift the two negative-helicity legs of $R$. For
example, after the first step we choose $\Shift {\hat 1} {\hK_{2,3}} $
as the shifted legs.  Similarly, when we perform a second insertion,
of $R_{n-2}$, we choose the intermediate $\hK$ momentum leg of the
last shift and the previously shifted $\hat{1}$ leg.

After this ``unwinding'' each resulting term can be expressed
schematically in the form
\begin{eqnarray}
&&\hspace*{-0.8cm} \Biggl( \prod_{r=2}^{j+1}
\frac{i\At_{3r}}
{K^2_r}
\Biggr)
A_{s}(1^-,\hK^-,(j+2)^+,\ldots,n^+) \,.
\label{eq:just_before_the_3_poin_id}
\end{eqnarray}
The product of three point gluon vertices contained inside the
brackets is equivalent to simply a tree amplitude divided by
$\spash{1}.{\hK_{2\ldots(j+1)}}^2$. Hence the recursion is solved as
\eqn{eq:just_before_the_3_poin_id} is written entirely in terms of
objects we know
\begin{eqnarray}
\frac{i\At_{j+1}(1^-,2^-,\ldots,(j+1)^+)}{\spash{1}.{\hK_{2\ldots(j+1)}}^2}
\Asdef(1^-,\hK_{2\ldots(j+1)}^-,(j+2)^+,\ldots,n^+)
\,.\nonumber
\end{eqnarray}

The complete unrenormalised scalar loop contribution is then given by
\begin{eqnarray}
A^{\textrm{scalar}}_{n}
= \cg \bigl[ \hat C_n + R^D_n \bigr] 
+ \frac{1}{3} A_n^{\NeqOne{\rm \ chiral}} + \cg\frac{2}{9} A_n^\tree\, ,
\end{eqnarray}
as in this case $\textrm{Inf}\;A_n=0$. The cut-completed contribution,
$\hat C_n$, previously calculated from unitarity techniques is given
in Ref.~\cite{Forde:2005hh} and
\begin{eqnarray}
R^D_n(1,2)
&=&\sum_{m=2}^{n-3}
\frac{i\At_{m}(1^-,2^-,\ldots ,m^+)}{\spa{1}.{\hK_{2\ldots m}}^2}
\Big(
\Asdef
(1^-,\hK_{2\ldots m}^-,(m+1)^+,\ldots ,n^+)
\nonumber\\
&&\hskip2cm \null
+ \Overlap_{n-m+2}
(1^-,\hK_{2\ldots
  m}^-,(m+1)^+,\ldots ,n^+)
\nonumber\\
&&\hskip2cm\null+\widehat{CR}_{n-m+2}
(1^-,\hK_{2\ldots
  m}^-,(m+1)^+,\ldots ,n^+)\Big)\, ,
\label{eq:complete_unwinding_result}
\end{eqnarray}
were in the recursion for $R_n^D$ both the overlap, $O_n$ pieces, and
the cut-completion $\widehat{CR}_n$ terms are included. Inserting the
known forms of these terms into \eqn{eq:complete_unwinding_result}
produces the result given in Ref.~\cite{Forde:2005hh}.

This ``unwinding'' technique also extends to other processes, so far
this has included all-multiplicity massive scalar
trees~\cite{FMassive} and the all-multiplicity MHV one-loop gluonic
QCD amplitude~\cite{Bootstrap}.  Hence the unitarity bootstrap
approach provides a bright new outlook on the calculation of
previously difficult to compute loop process needed to fully exploit
the promise of the LHC.

\end{document}